\documentstyle[psfig]{mn}

\newif\ifAMStwofonts

\ifoldfss
  \ifCUPmtlplainloaded \else
    \NewTextAlphabet{textbfit} {cmbxti10} {}
    \NewTextAlphabet{textbfss} {cmssbx10} {}
    \NewMathAlphabet{mathbfit} {cmbxti10} {} 
    \NewMathAlphabet{mathbfss} {cmssbx10} {} 
  \fi
  \ifAMStwofonts
    \ifCUPmtlplainloaded \else
      \NewSymbolFont{upmath} {eurm10}
      \NewSymbolFont{AMSa} {msam10}
      \NewMathSymbol{\upi}     {0}{upmath}{19}
      \NewMathSymbol{\umu}     {0}{upmath}{16}
      \NewMathSymbol{\upartial}{0}{upmath}{40}
      \NewMathSymbol{\leqslant}{3}{AMSa}{36}
      \NewMathSymbol{\geqslant}{3}{AMSa}{3E}

      \let\leq=\leqslant \let\le=\leqslant
       \let\ge=\geqslant
    \fi
  \fi
\fi 

\ifnfssone
  \newmathalphabet{\mathit}
  \addtoversion{normal}{\mathit}{cmr}{m}{it}
  \addtoversion{bold}{\mathit}{cmr}{bx}{it}
  \newmathalphabet{\mathbfit} 
  \addtoversion{normal}{\mathbfit}{cmr}{bx}{it}
  \addtoversion{bold}{\mathbfit}{cmr}{bx}{it}
  \newmathalphabet{\mathbfss} 
  \addtoversion{normal}{\mathbfss}{cmss}{bx}{n}
  \addtoversion{bold}{\mathbfss}{cmss}{bx}{n}
  \ifAMStwofonts
    \ifCUPmtlplainloaded \else
      \UseAMStwoboldmath
      \makeatletter
      \new@mathgroup\upmath@group
      \define@mathgroup\mv@normal\upmath@group{eur}{m}{n}
      \define@mathgroup\mv@bold\upmath@group{eur}{b}{n}
      \edef\UPM{\hexnumber\upmath@group}
      \new@mathgroup\amsa@group
      \define@mathgroup\mv@normal\amsa@group{msa}{m}{n}
      \define@mathgroup\mv@bold\amsa@group{msa}{m}{n}
      \edef\AMSa{\hexnumber\amsa@group}
      \makeatother
      \mathchardef\upi="0\UPM19
      \mathchardef\umu="0\UPM16
      \mathchardef\upartial="0\UPM40
      \mathchardef\leqslant="3\AMSa36
      \mathchardef\geqslant="3\AMSa3E

      \let\leq=\leqslant \let\le=\leqslant
       \let\ge=\geqslant
    \fi
  \fi
\fi 

\ifnfsstwo
  \DeclareMathAlphabet{\mathbfit}{OT1}{cmr}{bx}{it}
  \SetMathAlphabet\mathbfit{bold}{OT1}{cmr}{bx}{it}
  \DeclareMathAlphabet{\mathbfss}{OT1}{cmss}{bx}{n}
  \SetMathAlphabet\mathbfss{bold}{OT1}{cmss}{bx}{n}
  \ifAMStwofonts
    \ifCUPmtlplainloaded \else
      \DeclareSymbolFont{UPM}{U}{eur}{m}{n}
      \SetSymbolFont{UPM}{bold}{U}{eur}{b}{n}
      \DeclareSymbolFont{AMSa}{U}{msa}{m}{n}
      \DeclareMathSymbol{\upi}{0}{UPM}{"19}
      \DeclareMathSymbol{\umu}{0}{UPM}{"16}
      \DeclareMathSymbol{\upartial}{0}{UPM}{"40}
      \DeclareMathSymbol{\leqslant}{3}{AMSa}{"36}
      \DeclareMathSymbol{\geqslant}{3}{AMSa}{"3E}

      \let\leq=\leqslant \let\le=\leqslant
       \let\ge=\geqslant
    \fi
  \fi
\fi 

\ifCUPmtlplainloaded \else
  \ifAMStwofonts \else 
    \def\upi{\pi}
    \def\umu{\mu}
    \def\upartial{\partial}
  \fi
\fi
\def\gs{\mathrel{\raise1.16pt\hbox{$>$}\kern-7.0pt 
\lower3.06pt\hbox{{$\scriptstyle \sim$}}}}         
\def\ls{\mathrel{\raise1.16pt\hbox{$<$}\kern-7.0pt 
\lower3.06pt\hbox{{$\scriptstyle \sim$}}}}         

\title{Shapelets: I. A Method for Image Analysis}
\author[A. Refregier]{Alexandre Refregier\\
Institute of Astronomy, Madingley Road, Cambridge CB3 OHA, UK; 
ar@ast.cam.ac.uk}

\date{Accepted ---. Received ---; in original form ---.}

\pagerange{\pageref{firstpage}--\pageref{lastpage}}
\pubyear{2001}

\begin{document}

\maketitle

\label{firstpage}

\begin{abstract}
We present a new method for the analysis of images, a fundamental task
in observational astronomy. It is based on the linear decomposition of
each object in the image into a series of localised basis functions of
different shapes, which we call `Shapelets'. A particularly useful set
of complete and orthonormal shapelets is that consisting of weighted
Hermite polynomials, which correspond to perturbations around a
circular gaussian. They are also the eigenstates of the 2-dimensional
Quantum Harmonic Oscillator, and thus allow us to use the powerful
formalism developed for this problem. Among their remarkable
properties, they are invariant under Fourier transforms (up to a
rescaling), leading to an analytic form for convolutions. The
generator of linear transformations such as translations, rotations,
shears and dilatations can be written as simple combinations of
raising and lowering operators. We derive analytic expressions for
practical quantities, such as the centroid (astrometry), flux
(photometry) and radius of the object, in terms of its shapelet
coefficients. We also construct polar basis functions which are
eigenstates of the angular momentum operator, and thus have simple
properties under rotations.  As an example, we apply the method to
Hubble Space Telescope images, and show that the small number of
shapelet coefficients required to represent galaxy images lead to
compression factors of about 40 to 90. We discuss applications of
shapelets for the archival of large photometric surveys, for weak and
strong gravitational lensing and for image deprojection.
\end{abstract}

\begin{keywords}
methods: data analysis, analytical; techniques: image processing,
surveys, gravitational lensing
\end{keywords}

\section{Introduction}
\label{intro}
Image analysis is a fundamental task in observational astronomy.  For
instance, new techniques, such as weak gravitational lensing (see
reviews by Mellier 1999; Bartelmann \& Schneider 2000), microlensing
(Mao 1999) and the search for supernovae (Riess 2000; Perlmutter et
al. 1998), have great scientific promise, but require high precision
image processing and analysis. As a result, a number of sophisticated
data analysis packages (eg. FOCAS in IRAF, Jarvis \& Tyson 1986;
SExtractor, Bertin \& Arnouts 1996, etc) and techniques (eg. wavelet
analysis, see review by Stark, Murtagh \& Bijaoui 1998; image
subtraction, Alard \& Lupton 1998; shear measurement, Kaiser, Squires
\& Broadhurst 1995, Kaiser 2000, and Kuijken 2000) have been
developed.

In this paper, we present a new method for image analysis. It is based
on the linear decomposition of each object into a series of localised
basis functions with different shapes, which we call `Shapelets'. As a
basis set we choose weighted hermite polynomials. They correspond to
perturbations about a circular gaussian, and, in their asymptotic
form, to the Edgeworth expansion in several dimensions. They are also
the eigenstates of the 2-dimensional Quantum Harmonic Oscillator
(QHO), and thus allow us to use the powerful formalism developed for
this problem. They have remarkable properties. In particular, they are
(up to a rescaling) invariant under Fourier transforms and thus yield
a simple analytical form for convolutions. We derive a number of
practical tools which can use used to compute the characteristics of
the object (centroid, flux, radius, etc) from its shapelet
coefficients. Our method differs from the wavelet transform which
decomposes the image into a sum of basis functions of different scales
but with a set shape. In our method, the image is decomposed into a
collection of compact disjoint objects of arbitrary shapes and is thus
particularly adapted to astronomical images.

As an example, we show how images of galaxies observed with the Hubble
Space Telescope (HST) can be represented and strongly compressed using
shapelets. We also discuss several applications of shapelets, such as
archival of large photometric catalogues, gravitational lensing and
image de-projection. A precise method to measure the shear induced by
weak lensing on galaxy images is presented in an adjoining paper
(Refregier \& Bacon 2001, Paper II). The application of Shapelets to
interferometric images will be presented in Chang \& Refregier (2001).
The analytical results derived in this paper may also be useful for
any application using the Edgeworth expansion, such as, for instance,
the study of the growth of cosmological perturbations (Juskiewicz et
al. 1995 and reference therein).

This paper is organised as follows. In \S\ref{one_d}, we describe the
main properties of 1-dimensional shapelets and discuss their
connection to the QHO. In \S\ref{two_d_cartesian}, we show how
2-dimensional shapelets can be formed and derive a number of practical
analytical results. In \S\ref{convolution}, we discuss how the
shapelet states behave under convolutions. In \S\ref{two_d_polar}, we
derive polar shapelets from the cartesian basis functions and describe
some of their properties. In \S\ref{applications}, we discuss several
direct applications of shapelets. Our conclusions are presented in
\S\ref{conclusion}.

\section{One-Dimensional Shapelets}
\label{one_d}

\subsection{Definitions}
\begin{figure}
\psfig{figure=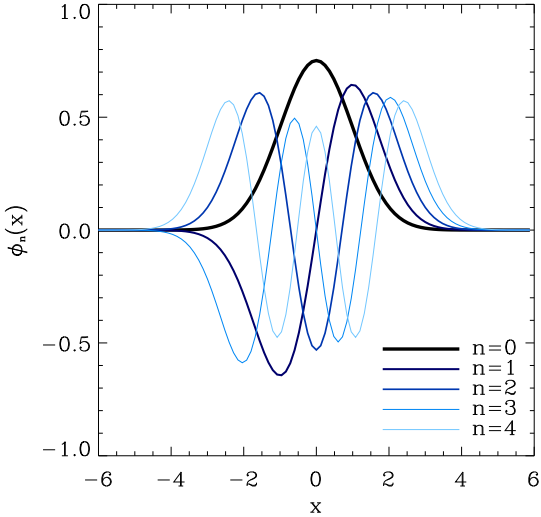,width=80mm} 
\caption{First few 1-dimensional basis functions $\phi_{n}(x)$.}
\label{fig:hermite}
\end{figure}

We first consider the description of a localised object in 1-dimension.
For this purpose, we first define the dimensionless basis functions
\begin{equation}
\phi_{n}(x) \equiv \left[ 2^{n}  \pi^{\frac{1}{2}}
  n! \right]^{-\frac{1}{2}} H_{n}(x) e^{-\frac{x^2}{2}}
\end{equation}
where $n$ is a non-negative integer and $H_{n}(x)$ is a hermite
polynomial of order $n$. These functions are orthonormal in the sense
that
\begin{equation}
\int_{-\infty}^{\infty} dx~\phi_{n}(x) \phi_{m}(x) = \delta_{nm},
\end{equation}
where $\delta_{mn}$ is the Kronecker delta symbol. The first few
functions are plotted on figure~\ref{fig:hermite}. These functions,
which we call `Shapelets', can be thought of as shape perturbations around
the gaussian $\phi_{0}(x)$,

To describe an object in practice, we use the dimensional basis
functions
\begin{equation}
B_{n}(x;\beta) \equiv \beta^{-\frac{1}{2}} \phi_{n}(\beta^{-1}x),
\end{equation}
where $\beta$ is a characteristic scale, which is typically chosen to
be close to the size of the object. These functions are also
orthonormal, i.e.
\begin{equation}
\label{eq:orthonorm}
\int_{-\infty}^{\infty} dx~B_{n}(x;\beta) B_{m}(x;\beta) = \delta_{nm}.
\end{equation}

This infinite set of functions forms a complete basis for smooth and
integrable functions. Thus, a (sufficiently well behaved) object
profile $f(x)$ can be expanded as
\begin{equation}
\label{eq:decompose}
f(x) = \sum_{n=0}^{\infty} f_{n} B_{n}(x;\beta).
\end{equation}
From the orthonormality condition (Eq.~[\ref{eq:orthonorm}]), the
shapelet coefficients are given by
\begin{equation}
f_{n} = \int_{-\infty}^{\infty} dx~f(x)B_{n}(x;\beta).
\end{equation}
In practice, the series of Equation~(\ref{eq:decompose}) will converge
quickly if the object $f(x)$ is sufficiently localised, and if $\beta$
and the origin $x=0$ are not too different from the size and location
of the object. This series representation is referred to as the
Gram-Charlier series, or, in its asymptotic form, as the Edgeworth
expansion (see eg. Juiszkiewicz 1995 and reference therein).

\subsection{Fourier Transform}
These basis functions have a number of useful properties. Let us first
consider their Fourier transform, which, for an arbitrary function
$f(x)$, is defined as
\begin{eqnarray}
\tilde{f}(k) & = & (2\pi)^{-\frac{1}{2}} \int_{-\infty}^{\infty} 
  dx f(x) e^{ikx}, \nonumber \\
f(x) & = & (2\pi)^{-\frac{1}{2}} \int_{-\infty}^{\infty} 
  dk \tilde{f}(k) e^{-ikx}.
\end{eqnarray}
With these conventions, the Fourier transform of the dimensionless
basis function $\phi_{n}(\xi)$ is
\begin{equation}
\label{eq:phi_tilde}
\widetilde{\phi}_{n}(\kappa) = i^{n} \phi_{n}(\kappa),
\end{equation}
Thus, up to a phase factor, the dimensionless basis functions are
invariant under Fourier transforms. This very useful property can be
understood in physical terms from the analogy with the quantum
harmonic oscillator (see \S\ref{qho}).

The Fourier transform of the dimensional basis function
$B_{n}(x;\beta)$ is given by 
\begin{equation}
\label{eq:B_tilde}
\widetilde{B}_{n}(k;\beta) = i^{n} B_{n}(k;\beta^{-1}).
\end{equation}
Thus, the Fourier transform acts on the basis functions with an
unsurprising change of scale $\beta \rightarrow \beta^{-1}$.

\subsection{Analogy with the Quantum Harmonic Oscillator}
\label{qho}
As we now discuss, the above basis functions are the eigenstates of
the Quantum Harmonic Oscillator (QHO), which
allows us to exploit the readily available formalism developed for
this problem. Let us consider a QHO with mass $m$ and natural
frequency $\omega$. If distances are measured in units of
$\sqrt{\frac{\hbar}{m\omega}}$ and energies in units of $\hbar
\omega$, the Hamiltonian for this system is
\begin{equation}
\label{eq:qho_hamiltonian}
\hat{H}=\frac{1}{2} \left[ \hat{x}^{2} + \hat{p}^{2} \right]
\end{equation}
where $\hat{x}$ and $\hat{p}$ are the position and momentum operators
respectively. In the $x$-representation, they are given by
\begin{equation}
\label{eq:xp}
\hat{x}=x,~~~\hat{p}=\frac{1}{i} \frac{\partial}{\partial x},
\end{equation}
and commute as $[ \hat{x}, \hat{p} ] = i$. As is well known, the basis
functions $\phi_{n}(x)$ are the eigenfunctions of the Hamiltonian,
with
\begin{equation}
\hat{H} \phi_{n} = \left( n+\frac{1}{2} \right) \phi_{n}.
\end{equation}
Clearly, $\hat{H}$ is symmetric under a permutation of $\hat{x}$ and
$\hat{p}$ (see Eq.~[\ref{eq:qho_hamiltonian}]), thus explaining the
invariance of $\phi_{n}$ under Fourier transforms
(Eq.~[\ref{eq:phi_tilde}]).

Of particular practical interest are the lowering and raising
operators, which are defined as
\begin{equation}
\label{eq:a_adagger}
\hat{a} \equiv \frac{1}{\sqrt{2}} \left(\hat{x} + i \hat{p} \right),~~~
\hat{a}^{\dagger} \equiv \frac{1}{\sqrt{2}} \left(\hat{x} - i \hat{p}
\right),
\end{equation}
where $^{\dagger}$ is the Hermitian conjugate. They commute as
$[\hat{a},\hat{a}^{\dagger}] = 1$, and act on the basis functions as
\begin{equation}
\label{eq:a_action}
\hat{a} \phi_{n} = \sqrt{n} \phi_{n-1},~~~ 
\hat{a}^{\dagger} \phi_{n} = \sqrt{n+1} \phi_{n+1}.
\end{equation}
The Hamiltonian can then be rewritten as $\hat{H} = \hat{N} +
\frac{1}{2}$, where the number operator $\hat{N} \equiv
\hat{a}^{\dagger} \hat{a}$ has the property that
\begin{equation}
\hat{N} \phi_{n} = n \phi_{n}.
\end{equation}
When convenient, we will use the bra-ket notation of quantum
mechanics. For instance, the $n^{\rm th}$ state is written as
$| n \rangle$ and has an $x$-representation given by
$\langle x | n \rangle = \phi_{n}(x)$.

The dimensional basis functions are the eigenfunctions of the
Hamiltonian 
\begin{equation}
\hat{H}_{\beta} = \frac{1}{2} \left[ \beta^{-2} \hat{x}^{2}
  + \beta^{2} \hat{p}^{2} \right].
\end{equation}
The eigenstates are labeled as $|n;\beta \rangle$ and obey
$\hat{H}_{\beta} |n;\beta \rangle = \left( n+\frac{1}{2} \right)
|n;\beta \rangle$. The dimensional basis functions are then given by
$B_{n}(x;\beta) = \langle x | n; \beta \rangle$.

\subsection{Further Properties}
The Hermite basis functions have a number of further convenient
properties which we will need later and summarise here.

We first notice, by inspecting Figure~\ref{fig:hermite}, that the
basis functions $B_{n}(x,\beta)$ acquire both a larger extent and
smaller scale oscillations when the order $n$ is increased, keeping
$\beta$ constant. This can be described more precisely by considering
the characteristic radius $\theta_{\rm max}(\beta,n)$ of a basis
function, defined by $\theta_{\rm max}^{2}(\beta,n) \equiv \langle n;
\beta | \hat{x}^2 | n;\beta \rangle$. As is well known from Quantum
Mechanics and can easily derived using Equation~(\ref{eq:a_adagger}),
this rms radius is given by $\theta_{\rm max}(\beta,n) = \beta \left(
n+\frac{1}{2} \right)^{\frac{1}{2}}$. Similarly, the characteristic
size $\theta_{\rm min}(\beta,n)$ of the small scale (oscillatory)
features in a basis function of order $n$ is defined as the rms
inverse radius in Fourier space, i.e. by $\theta_{\rm
min}^{-2}(\beta,n) \equiv \langle n; \beta | \hat{p}^2 | n;\beta
\rangle$. As can again be verified using raising and lower operators,
the radius is given by $ \theta_{\rm min}(\beta,n) = \beta \left(
n+\frac{1}{2} \right)^{-\frac{1}{2}}$.  Thus a decomposition which
includes modes from $n=0$ to $n_{\rm max}$ can represent features with
scales ranging between the two limits $\theta_{\rm min}(\beta,n_{\rm
max})$ and $\theta_{\rm max}(\beta,n_{\rm max})$.  In
\S\ref{astrometry} below, we show how these scales can be used to fine
a good choice of $\beta$ for an object.
\label{th_min_max_1d}.

Another important property relates to the rescaling of a shapelet
function. Let us for instance consider a function $f(x)=\sum_{n} f_{n}
B_{n}(x;\beta)$, which has been decomposed into shapelets of
scale $\beta$.  It can be sometimes convenient to express it in terms
of basis functions with a different scale $\beta'$, as $f(x)=\sum_{n}
f_{n}' B_{n}(x;\beta')$. The relation between the coefficients $f_{n}$
and $f_{n}'$ is derived in Appendix~\ref{rescaling} and involves the
overlap matrix $\langle n;\beta | n',\beta' \rangle$, whose analytic
form is given by Equation~(\ref{eq:beta1_beta2}).

Finally, we note that the basis functions obey the integral property
\begin{equation}
\label{eq:1n}
<1|n;\beta> \equiv \int_{-\infty}^{\infty} dx~B_{n}(x;\beta) = 
\left[ 2^{1-n} \pi^{\frac{1}{2}} \beta \right]^{\frac{1}{2}}
\left( \begin{array}{c} n \\ n/2 \end{array} \right)^{\frac{1}{2}},
\end{equation}
for $n$ even (the integral vanishes otherwise), where the parenthesis
denotes the binomial coefficient and a convenient shorthand notation
was used on the left-hand side. This can be derived using the
generating function of Hermite polynomials (eg. Arfken 1985).

\section{Two-Dimensional Cartesian Shapelets}
\label{two_d_cartesian}
In this section, we construct 2-dimensional shapelets by taking
products of the 1-dimensional shapelets described above.  We then
study the properties of the resulting `Cartesian' basis functions, and
derive a number of analytical results which are useful in practice.

\subsection{Definitions}
\label{2d_cartesian_def}
Basis functions for 2-dimensional objects can be constructed by taking
the tensor product of two 1-dimensional basis functions. We thus
define the dimensionless functions
\begin{equation}
\label{eq:phi_n1n2}
\phi_{{\mathbf n}}({\mathbf x}) \equiv \phi_{n_{1}}(x_{1})
\phi_{n_{2}}(x_{2}),
\end{equation}
where ${\mathbf x}=(x_{1},x_{2})$ and ${\mathbf
n}=(n_{1},n_{2})$. Dimensional basis functions are defined as.
\begin{equation}
B_{{\mathbf n}}({\mathbf x};\beta) \equiv \beta^{-1} 
  \phi_{{\mathbf n}}(\beta^{-1}{\mathbf x}).
\end{equation}
These 2-dimensional shapelets are again orthonormal, in the sense that
\begin{equation}
\int d^{2}x~B_{{\mathbf n}}({\mathbf x};\beta)
  B_{{\mathbf m}}({\mathbf x};\beta) = \delta_{n_{1}m_{1}}
  \delta_{n_{2}m_{2}}.
\end{equation}
The functions $\phi_{{\mathbf n}}({\mathbf x})$ are eigenstates of the
2-dimensional QHO whose Hamiltonian is
\begin{equation}
\label{eq:hamiltonian_2d}
\hat{H} = \frac{1}{2} \left[ \hat{x}_{1}^{2} + 
  \hat{x}_{2}^{2} + \hat{p}_{1}^{2} + \hat{p}_{2}^{2} \right],
\end{equation}
where $\hat{x}_{i}$ and $\hat{p}_{i}$ are the position and momentum
operators for each dimension.

\begin{figure}
\psfig{figure=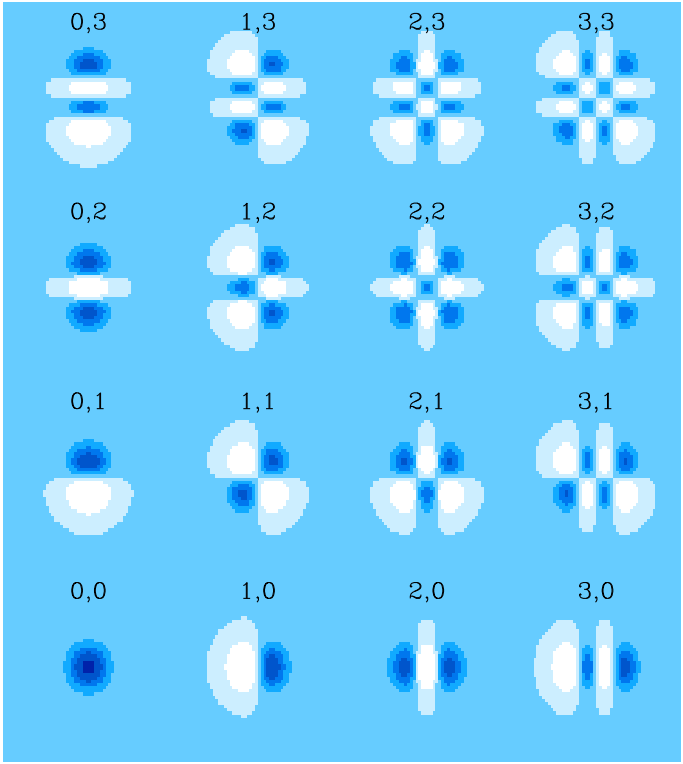,width=90mm} 
\caption{First few 2-dimensional Cartesian basis functions
$\phi_{n_{1},n_{2}}$. The dark and light regions correspond
to positive and negative values, respectively.}
\label{fig:phi}
\end{figure}

\begin{figure}
\psfig{figure=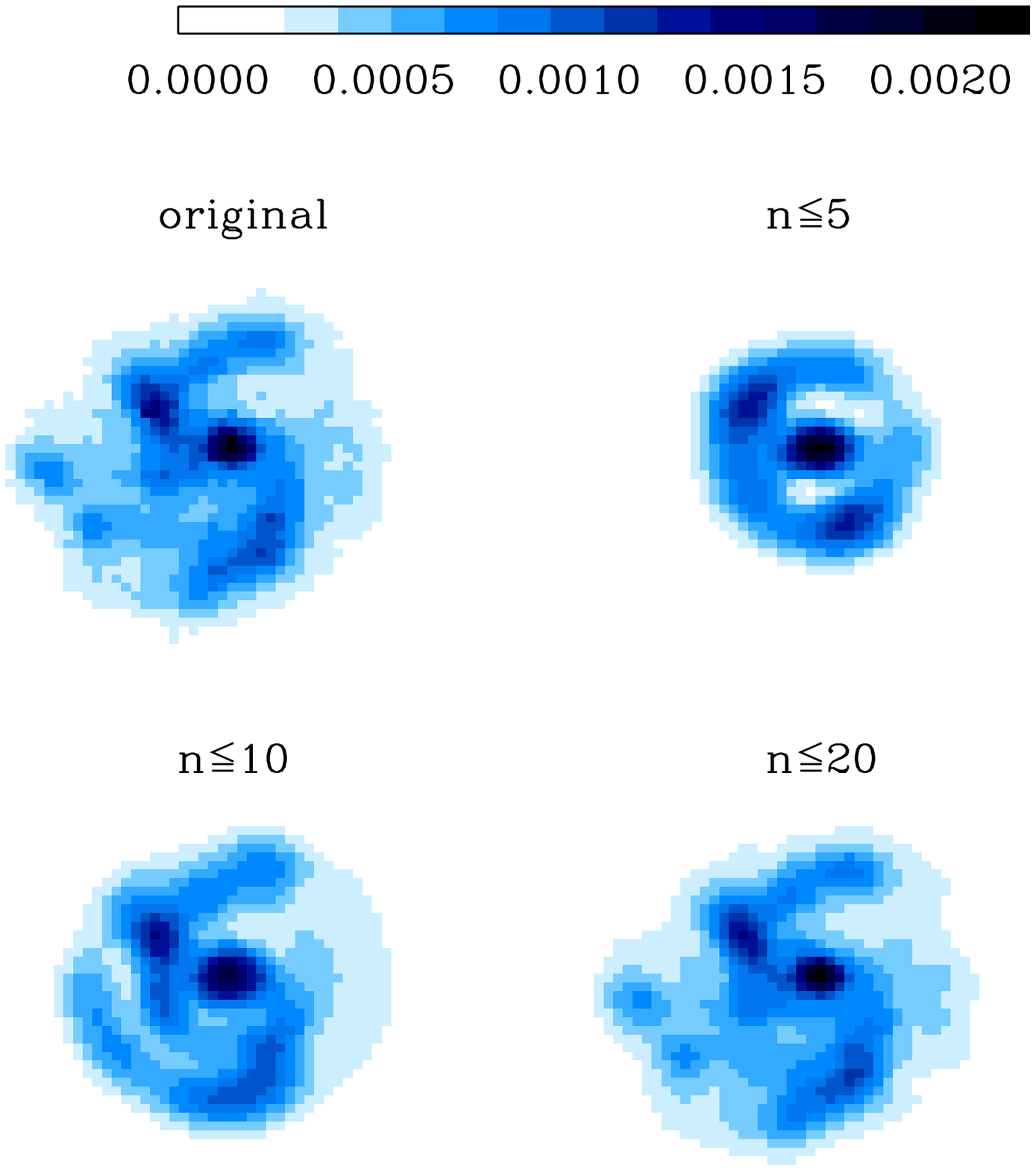,width=90mm} 
\caption{Decomposition of a galaxy image found in the HDF. The
original $60\times60$ pixel HST image (upper left-hand panel) can be
compared with the reconstructed images with different maximum order
$n=n_{1}+n_{2}$. The shapelet scale is chosen to be $\beta=4$
pixels. The lower right-hand panel ($n \leq 20$) is virtually
indistinguishable from the initial image.}
\label{fig:hst}
\end{figure}

\begin{figure}
\psfig{figure=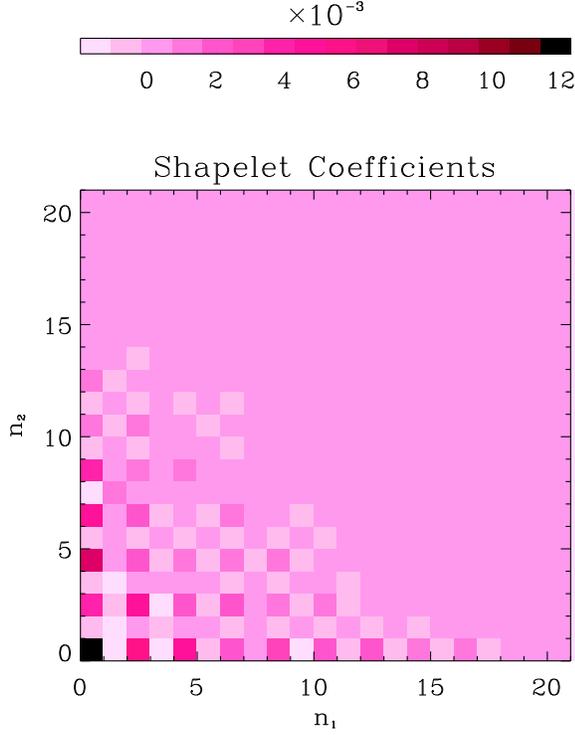,width=90mm} 
\caption{Shapelet coefficients for the image decomposition of
the previous figure. Since the coefficient array is sparse, the images
can be reconstructed from the few first largest coefficients.}
\label{fig:hst_coef}
\end{figure}

\begin{figure}
\psfig{figure=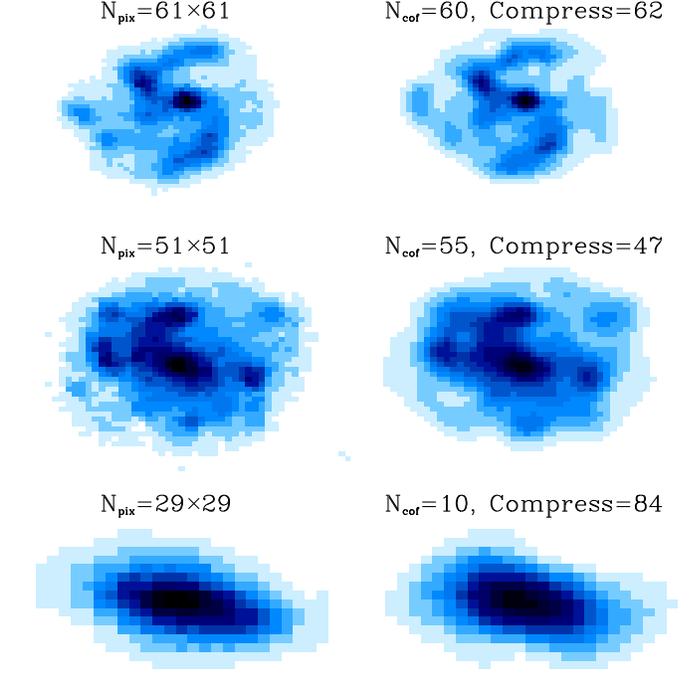,width=90mm} 
\caption{Reconstruction and compression of three HST galaxy images
using shapelets. The left-hand column shows the orginal images
extracted from the HDF and list $N_{\rm pix}$ their size in
pixels. The right-hand column shows their reconstructed image from the
$N_{\rm cof}$ largest coefficients (in absolute value) of their
shapelet decomposition. Because the coefficient matrix is typically
sparse, a large compression factor $N_{\rm pix}/N_{\rm cof}$ is
achieved. The shapelet scale was chosen to be $\beta=4$ pixels in all 3
cases.}
\label{fig:compress}
\end{figure}

The first few 2-dimensional shapelets are shown on
Figure~\ref{fig:phi}. Again, they can be thought of as perturbations
around the 2-dimensional Gaussian $\phi_{00}$. These basis functions
form a complete orthonormal basis for smooth, integrable functions of
two variables. A (well behaved) 2-dimensional function $f({\mathbf
x})$, such as the image of an object, can thus be decomposed as
\begin{equation}
\label{eq:decompose_2d}
f({\mathbf x}) = \sum_{n_1,n_2=0}^{\infty} f_{\mathbf n} 
  B_{\mathbf n}({\mathbf x};\beta),
\end{equation}
where the shapelet coefficients are given by
\begin{equation}
\label{eq:coefs_2d}
f_{\mathbf n} = \int d^{2}x f({\mathbf x}) B_{\mathbf n}({\mathbf
x};\beta)
\end{equation}
Figure~\ref{fig:hst} show how an image observed with HST can be
decomposed and reconstructed using shapelets. The resulting
distribution of the coefficients is shown on
Figure~\ref{fig:hst_coef}. More examples can be found on
Figure~\ref{fig:compress}.  These examples and associated applications
will be discussed in detail in \S\ref{applications}.

Of practical interest, is the choice of an appropriate shapelet scale
$\beta$ and maximum order $n_{\rm max}$ for the faithful and
efficient decomposition of a given image. Using arguments similar to
those of \S\ref{th_min_max_1d}, it is easy to show that a decomposition
in 2-dimensions which include shapelets of scale $\beta$ with order
ranging from $n_1+n_2=0$ to $n_{\rm max}$ can only describe features
with scales between the two limits
\begin{equation}
\theta_{\rm min} \approx \beta 
  \left( n_{\rm max}+1 \right)^{-\frac{1}{2}}, ~~
\theta_{\rm max} \approx \beta \left( n_{\rm max}+1 \right)^{\frac{1}{2}}.
\end{equation}
Thus, if the function has features with scales ranging from
$\theta_{\rm max}$ (eg. the size of the object or that of the image)
and $\theta_{\rm min}$ (eg. the pixel size, or the size of a smoothing
kernel), a good choice of $\beta$ and $n_{\rm max}$ will be
\begin{equation}
\beta \approx (\theta_{\rm min} \theta_{\rm max})^{\frac{1}{2}},~~ n_{\rm max}
\approx \frac{\theta_{\rm max}}{\theta_{\rm min}} - 1.
\end{equation}
In practice, this provides a good first guess, which can be refined
using a few iterations (see \S\ref{astrometry}).
\label{th_min_max}

\subsection{Photometry and Astrometry}
\label{astrometry}
The most basic quantities to measure for an object image are its
total flux (photometry), centroid position (astrometry) and size. Let
us first decompose the intensity $f({\mathbf x})$ of the object into
shapelet coefficients $f_{\mathbf n}=\langle {\mathbf n};\beta|f\rangle$ as in
Equation~(\ref{eq:decompose_2d}).

Using the integral property of Equation~(\ref{eq:1n}), it is then easy
to show that the total flux $F\equiv \int d^{2}x f({\mathbf x})$ of
the object is
\begin{equation}
F = \pi^{\frac{1}{2}} \beta \sum_{n1,n2}^{\rm even} 
2^{\frac{1}{2}(2-n_{1}-n_{2})} 
\left( \begin{array}{c} n_{1} \\ n_{1}/2 \end{array} \right)^{\frac{1}{2}}
\left( \begin{array}{c} n_{2} \\ n_{2}/2 \end{array}
\right)^{\frac{1}{2}} f_{n_1 n_2},
\end{equation}
where the sum is over even values of $n_{1}$ and $n_{2}$.

Using Equations~(\ref{eq:1n}) and (\ref{eq:a_adagger}), one can also
show that the centroid of the object $x_{i}^{f} \equiv \int d^{2}x
x_{i} f({\mathbf x})/F$ is given by
\begin{eqnarray}
x_{1}^{f} & = & \pi^{\frac{1}{2}} \beta^{2} F^{-1} \sum_{n1}^{{\rm odd}}
\sum_{n2}^{{\rm even}} (n_{1}+1)^{\frac{1}{2}}  2^{\frac{1}{2}(2-n_{1}-n_{2})} 
  \nonumber \\ 
 & & \times
\left( \begin{array}{c} n_{1}+1 \\ (n_{1}+1)/2 \end{array} 
  \right)^{\frac{1}{2}}
\left( \begin{array}{c} n_{2} \\ n_{2}/2 \end{array} \right)^{\frac{1}{2}}
f_{n_1 n_2}, 
\end{eqnarray}  
and similarly for $x_{2}^{f}$. 

Similarly, the rms radius $r_{f}$ defined by $r_{f}^{2} \equiv \int
d^{2}x x^{2} f({\mathbf x})/F$ is given by
\begin{eqnarray}
r_{f}^{2} & = & \pi^{\frac{1}{2}} \beta^{3} F^{-1} \sum_{n1,n2}^{\rm even} 
2^{\frac{1}{2}(4-n_{1}-n_{2})} \left( 1+n_{1}+n_{2} \right) \nonumber \\
 &  & \times
\left( \begin{array}{c} n_{1} \\ n_{1}/2 \end{array} \right)^{\frac{1}{2}}
\left( \begin{array}{c} n_{2} \\ n_{2}/2 \end{array} \right)^{\frac{1}{2}} 
f_{n_1 n_2},
\end{eqnarray}
These expressions can be used, by iteration, to find the optimal centre
and scale of the basis functions.

\subsection{Coordinate Transformations}
\label{transformations}
Let us consider a general coordinate transformation of the form
${\mathbf x} \rightarrow {\mathbf x}'=(1+{\mathbf \Psi}) {\mathbf x} +
{\mathbf \epsilon}$, where ${\mathbf \Psi}$ is a $2\times2$ matrix,
${\mathbf \epsilon}=(\epsilon_{1},\epsilon_{2})$ is a small
displacement. Such a transformation can arise for instance from a
translation, rotation or from the action of gravitational lensing. We
assume that the transformation matrix ${\mathbf \Psi}$ and the
displacement ${\mathbf \epsilon}$ are small and constant across the
object. We parametrise the matrix ${\mathbf \Psi}$ following the
gravitational lensing conventions as
\begin{equation}
\label{eq:psi_params}
{\mathbf \Psi} =
 \left( \begin{array}{cc}
\kappa +\gamma_{1} & \gamma_{2} - \rho \\
\gamma_{2} + \rho & \kappa - \gamma_{1} \\
\end{array} \right),
\end{equation}
where $\rho$ describes rotations and the convergence $\kappa$
describes overall dilatations and contractions. The shear $\gamma_{1}$
($\gamma_{2}$) describes stretches and compressions along (at
$45^{\circ}$ from) the x-axis. The displacements $\epsilon_1$ and
$\epsilon_2$ correspond to translations along the $x$ and $y$-axis,
respectively.

Under this transformation, the intensity $f({\mathbf x})$ of an object
becomes
\begin{equation}
f'({\mathbf x}') = f({\mathbf x}({\mathbf x}')) \simeq
f({\mathbf x}'-{\mathbf \Psi}{\mathbf x}'-{\mathbf \epsilon}).
\end{equation}
Since we are now considering infinitesimal transformations, we can
Taylor expand this expression and only keep the terms which are first
order in $\Psi$. After using Equations~(\ref{eq:xp}) 
and (\ref{eq:a_adagger}),
we find
\begin{equation}
f' \simeq (1 + \rho \hat{R}+ \kappa \hat{K} + \gamma_{j} \hat{S}_{j}
+ \epsilon_{i} \hat{T}_{i}) f,
\end{equation}
where $\hat{R}$, $\hat{K}$, $\hat{S}_{i}$ and $\hat{T}_{i}$ are the
operators generating rotation, convergence, shears and translations,
respectively, and where we have used the Einstein summation convention.
The generators are given by
\begin{eqnarray}
\label{eq:generators}
\hat{R} & = & -i \left( \hat{x}_{1} \hat{p}_{2} - \hat{x}_{2}
\hat{p}_{1} \right) =
\hat{a}_1 \hat{a}_2^{\dagger} - \hat{a}_1^{\dagger}
  \hat{a}_2 \nonumber \\ 
\hat{K} & = & -i \left( \hat{x}_{1} \hat{p}_{1} + \hat{x}_{2}
\hat{p}_{2} \right) =
 1 + \frac{1}{2} \left( \hat{a}_{1}^{\dagger 2} + 
 \hat{a}_{2}^{\dagger 2} -\hat{a}_{1}^{2} - \hat{a}_{2}^{2} \right)
  \nonumber \\ 
\hat{S}_{1} & = & -i \left( \hat{x}_{1} \hat{p}_{1} - \hat{x}_{2}
\hat{p}_{2} \right) =
\frac{1}{2} \left( \hat{a}_{1}^{\dagger 2} -
  \hat{a}_{2}^{\dagger 2} - \hat{a}_{1}^{2} + \hat{a}_{2}^{2} \right)
  \nonumber \\ 
\hat{S}_{2} & = & -i \left( \hat{x}_{1} \hat{p}_{2} + \hat{x}_{2}
\hat{p}_{1} \right) =
\hat{a}_1^{\dagger} \hat{a}_2^{\dagger} - \hat{a}_1 \hat{a}_2
\nonumber \\
\hat{T}_{j} & = & - i \hat{p}_{j} = \frac{1}{\sqrt{2}} (\hat{a}_{j}^{\dagger}
- \hat{a}_{j}),~~~j=1,2.
\end{eqnarray}
The rotation generator $\hat{R}$ is thus simply equal to the angular
momentum operator in 2-dimensions
\begin{equation}
\label{eq:l}
\hat{L}=\hat{x}_{1} \hat{p}_{2} - \hat{x}_{2} \hat{p}_{1}=
i \left (\hat{a}_1 \hat{a}_2^{\dagger} - \hat{a}_1^{\dagger} \hat{a}_2 \right),
\end{equation}
up to a factor of $-i$. Similarly, the
translation generator $\hat{T}_{i}$ is simply equal to the the linear
momentum operator $\hat{p}_i$, up to the same factor.

These expressions along with Equation~(\ref{eq:a_action}) make it easy
to compute the effect of these transformations on the basis functions
$B_{\mathbf n}$. For instance, the generator of translations along the
$x$-axis has a matrix representation $T_{1{\mathbf mn}} \equiv
\langle {\mathbf m} | \hat{T}_{1} | {\mathbf n} \rangle$ given by
\begin{equation}
T_{1{\mathbf mn}} = \frac{1}{\sqrt{2}} \left[ 
\sqrt{n_1} \delta_{m_1,n_1-1} 
- \sqrt{n_1+1} \delta_{m_1,n_1+1} \right] \delta_{m_2,n_2},
\end{equation}
and similarly for the other generators.

\begin{figure}
\psfig{figure=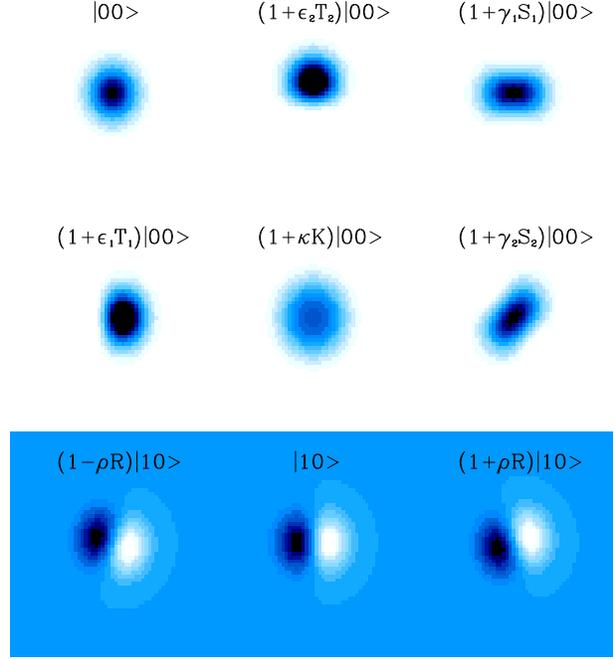,width=80mm}
\caption{Effect of coordinate transformations on the first two
shapelet states $|00\rangle$ and $|10\rangle$. All coordinate
transformations are considered: translations, rotations, convergence
and shear. They are parametrised by $\epsilon_{i}$, $\rho$, $\kappa$
and $\gamma_{i}$, respectively, which were all assigned a value of 0.3
for the purpose of this figure.  Their action on any state are
easily calculated using the raising and lowering operators
$\hat{a}_{i}^{\dagger}$ and $\hat{a}_{i}$. Clearly, the different
transformation generators act as
expected from their definition.}
\label{fig:gen}
\end{figure}

The meaning of the generators can be seen by studying their action on
the ground state. For instance, it is easy to see that under the
action of a shear $\gamma_{1}$, the ground state $|00\rangle$ becomes
\begin{equation}
|00\rangle'\simeq (1+\gamma_1 \hat{S}_1)|00\rangle = |00\rangle +
\frac{\gamma_1}{\sqrt{2}} \left[ |20\rangle - |20\rangle \right].
\end{equation}
The action of the different transformations on the ground states can
be calculated in the same way and are shown in
Figure~\ref{fig:gen}. Clearly, their action is as expected from their
definition. Since the ground state is circularly symmetric, the
rotation operator vanishes when applied to $|00\rangle$,
i.e. $\hat{R}|00\rangle = 0$. More instructively, we can consider the
effect of $\hat{R}$ on an asymmetric state like $|10\rangle$. It is
also shown in the bottom row of the figure. As expected the state
$|10\rangle$ is rotated counter-clockwise by the rotation operator.

Finite transformations can be produced by exponentiating the
generators. For instance, after a finite rotation by an angle $\rho$
the function $f$ becomes
\begin{equation}
\label{eq:rot_finite}
f' = e^{\rho \hat{R}} f = \left( \sum_{n=0}^{\infty} \frac{(\rho
\hat{R})^{n}}{n!} \right) f,
\end{equation}
and similarly for the shear and convergence. 

\subsection{Effect of Noise}
In this section, we study the uncertainty induced by noise on the
basis function decomposition, in the case of correlated and
uncorrelated background noise, and of Poisson noise. The observed
intensity of an object is
\begin{equation}
f'({\mathbf x}) = f({\mathbf x}) + n({\mathbf x}),
\end{equation}
where $f({\mathbf x})$ is the intrinsic intensity of the object and
$n({\mathbf x})$ is the noise.  The noise is taken to be unbiased, so
that $\langle n({\mathbf x}) \rangle = 0$ and is characterised by its
correlation function $\eta({\mathbf x},{\mathbf x}') \equiv \langle
n({\mathbf x}) n({\mathbf x}') \rangle$. Here the brackets refer to
an ensemble average over noise realisations.

The observed basis coefficients are then $f'_{{\mathbf k}} = \langle
{\mathbf k}, \beta | f'\rangle$ and are unbiased, i.e.
\begin{equation}
\langle f'_{{\mathbf k}} \rangle = f_{{\mathbf k}},
\end{equation}
where $f_{{\mathbf k}} = \langle {\mathbf k}, \beta | f \rangle$ are
the intrinsic coefficients. It is then easy to show that the
covariance matrix ${\rm cov}[f'_{{\mathbf k}},f'_{{\mathbf l}}] \equiv
\langle (f'_{{\mathbf k}} - \langle f'_{{\mathbf k}} \rangle
)(f'_{{\mathbf l}} - \langle f'_{{\mathbf l}} \rangle) \rangle$ for
the observed coefficients is given by
\begin{equation}
\label{eq:cov_general}
{\rm cov}[f'_{{\mathbf k}},f'_{{\mathbf l}}] = 
  \int d^{2}x \int d^{2}x' ~B_{k}({\mathbf x},\beta) 
  \eta({\mathbf x},{\mathbf x}') B_{l}({\mathbf x},\beta). 
\end{equation}

To be more specific, we first consider the case of homogeneous
background noise, as can be produced by sky or instrumental
backgrounds. If the background noise is uncorrelated, $\eta({\mathbf
x}) = \sigma_{n}^{2} \delta^{(2)}({\mathbf x})$, where $\sigma_{n}$ is
the {\it rms} noise. As a result, the covariance matrix reduces to
\begin{equation}
{\rm cov}[f'_{{\mathbf k}},f'_{{\mathbf l}}] = \sigma_{n}^{2}
\delta_{lk},
\end{equation}
where we have used the orthonormality of the basis functions
(Eq.~[\ref{eq:orthonorm}]). In this case, the covariance matrix is
thus diagonal, so that each coefficient is statistically
independent. Moreover, the diagonal elements are all equal.
Uncorrelated noise thus populates each coefficient equally, and is
thus ``white'' as in the case of Fourier transforms.

In the case of spatially correlated but homogeneous noise, the noise
correlation function is only a function of separation and can thus be
written as $\eta({\mathbf x} - {\mathbf x}')$. As a result,
Equation~(\ref{eq:cov_general}) reduces to the integral of a
convolution and can thus be written symbolically as
\begin{equation}
{\rm cov}[f'_{{\mathbf k}},f'_{{\mathbf l}}] = 
\langle {\mathbf k}, \beta | \eta * ({\mathbf l}, \beta) \rangle,
\end{equation}
in the notation of Equation~(\ref{eq:cnml_notation}). A convenient
way to evaluate this is to decompose $\eta({\mathbf x})$ itself into
basis functions and then to use the results of \S\ref{convolution}
below. Spatial correlations in the noise thus produces correlations in
the coefficients.

Another case of practical interest is that in which the noise is
dominated by Poisson shot noise. If the intensities are measured in
units of photon counts, the noise correlation function is then
$\eta({\mathbf x},{\mathbf x}') = f({\mathbf x}) \delta^{(2)}({\mathbf
x}-{\mathbf x}')$. As a result, the covariance matrix is
\begin{equation}
{\rm cov}[f'_{{\mathbf k}},f'_{{\mathbf l}}] = 
\sum_{{\mathbf m}} f_{{\mathbf m}}
  B^{(3)}_{{\mathbf k},{\mathbf l},{\mathbf m}},
\end{equation}
where $B^{(3)}_{{\mathbf k},{\mathbf l},{\mathbf
m}}(\beta,\beta,\beta)$ is the 3-product integral defined in
Equation~(\ref{eq:b3}) below, and which is evaluated analytically in
Paper II. In this case again, the covariance coefficient is made
non-diagonal by the noise correlation, but is easily calculable
analytically.

\section{Convolution}
\label{convolution}
We now show how shapelets behave under convolutions, an operation
which often occurs in practice (eg. under the action of PSF, seeing,
smoothing, etc). We start by considering convolution by a general
kernel in 1-dimensions, and then study the special case of smoothing
by a gaussian. Finally, we treat the 2-dimensional case, and
illustrate the results with the example of an HST galaxy image.

\subsection{Convolution in 1-Dimension}
\label{convolve}
Let us first consider the convolution of two arbitrary 1-dimensional
functions $f(x)$ and $g(x)$. Their convolution $h(x)$ can be written as
\begin{equation}
h(x) \equiv (f * g)(x) \equiv \int_{-\infty}^{\infty}
dx'~f(x-x')g(x')
\end{equation}
Each function can be decomposed into our basis functions with scales
$\alpha$, $\beta$ and $\gamma$. These scales are chosen to be most
convenient in each case. The coefficients are then $f_{n} \equiv \langle
n;\alpha | f \rangle$, $g_{n} \equiv \langle n;\beta | g \rangle$,
$h_{n} \equiv \langle n;\gamma | h \rangle$. Our aim is to find an expression
which relates $h_{n}$ to $f_{n}$ and $g_{n}$. Since convolution is a
bi-linear operation, this relation will be of the form
\begin{equation}
h_{n} = \sum_{m,l=0}^{\infty} C_{nml} f_{m} g_{l},
\end{equation}
where the convolution tensor can be written symbolically as
\begin{equation}
\label{eq:cnml_notation}
C_{nml}(\gamma,\alpha,\beta) \equiv \langle n;\gamma | (m;\alpha)*
(l;\beta) \rangle
\end{equation}
 and is a function of the scale lengths. Using the properties of the
basis functions under Fourier transforms (Eq.~[\ref{eq:B_tilde}]), it
is easy to show that the convolution tensor is given by
\begin{equation}
\label{eq:c_nml}
C_{nml}(\gamma,\alpha,\beta)=(2\pi)^{\frac{1}{2}} (-1)^{n} i^{n+m+l}
B^{(3)}_{nml}(\gamma^{-1},\alpha^{-1},\beta^{-1}),
\end{equation}
where the 3-product integral is $B^{(3)}_{nml}(a_1,a_2,a_3)$
is defined as
\begin{equation}
\label{eq:b3}
B^{(3)}_{lmn}(a_1,a_2,a_3) \equiv \int_{-\infty}^{\infty} dx~
  B_{l}(x,a_1) B_{m}(x,a_2) B_{n}(x,a_3).
\end{equation}
As we show in Paper II, this integral can be easily
evaluated analytically with a recurrence relation.

\subsection{Smoothing in 1-Dimension}
\label{smoothing_1d}
The special case consisting of smoothing by a gaussian is useful in
practice. In this case, we let
\begin{equation}
g(x) \equiv (2\pi)^{-\frac{1}{2}} \beta^{-1}
e^{-\frac{x^2}{2\beta^{2}}},
\end{equation}
which is normalised so that $\int dx~g(x)=1$. We can then write the
coefficients for the smoothed function $h(x)$ as
\begin{equation}
h_{n}=\sum_{m} G_{nm} f_{m},
\end{equation}
where $G_{nm}(\gamma,\alpha,\beta) = \sum_{l}
C_{nml}(\gamma,\alpha,\beta) g_{l}$ is the smoothing matrix. The
gaussian $g(x)$ can be thought as a (non-normalised) $n=0$ shapelet
state of amplitude $g_{0}=\langle 0;\beta | g \rangle$, so that
$G_{nm} = C_{nm0} g_{0}$. Using the generating function for Hermite
polynomials, one can show that, for the natural choice of
$\gamma^{2}=\alpha^{2}+\beta^{2}$, the smoothing matrix is given by
\begin{equation}
\label{eq:g_nm}
G_{nm} = 2^{\frac{n-m}{2}} 
\left( \frac{\omega}{\beta} \right)^{\frac{1}{2}}
\frac{\omega^{m}}{\beta^{n}\alpha^{m-n}}
\frac{(m!/n!)^{\frac{1}{2}}}
{\left( \frac{m-n}{2}\right)!},
\end{equation}
for $m-n\ge0$ and even ($G_{nm}$ vanishes otherwise), and where
$\omega^{-2} \equiv \alpha^{-2}+\beta^{-2}$.

\begin{figure}
\psfig{figure=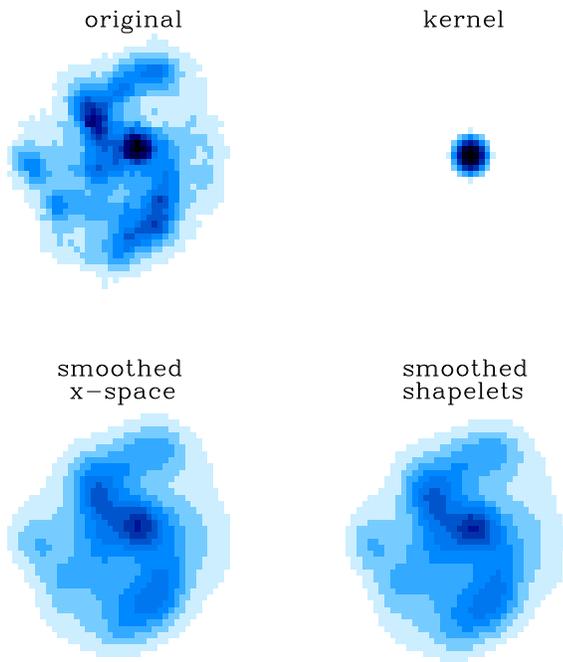,width=90mm}
\caption{Illustration of smoothing in Shapelet Space. The original
galaxy image ($61 \times 61$ pixels) of Figure~\ref{fig:hst} (shown
again in the upper-left panel) is smoothed with a gaussian kernel with
standard deviation $\beta=2$ pixels (upper-right panel).  The
resulting image smoothed using shapelets (lower-right panel) is
almost indistinguishable from that smoothed using direct
convolution in real-space (lower-left panel). In shapelet space,
smoothing is a simple matrix multiplication and can be very efficient
when the coefficient matrix is sparse, as is the case here (see
Figure ~\ref{fig:hst_coef}).}
\label{fig:smooth}
\end{figure}

\begin{figure}
\psfig{figure=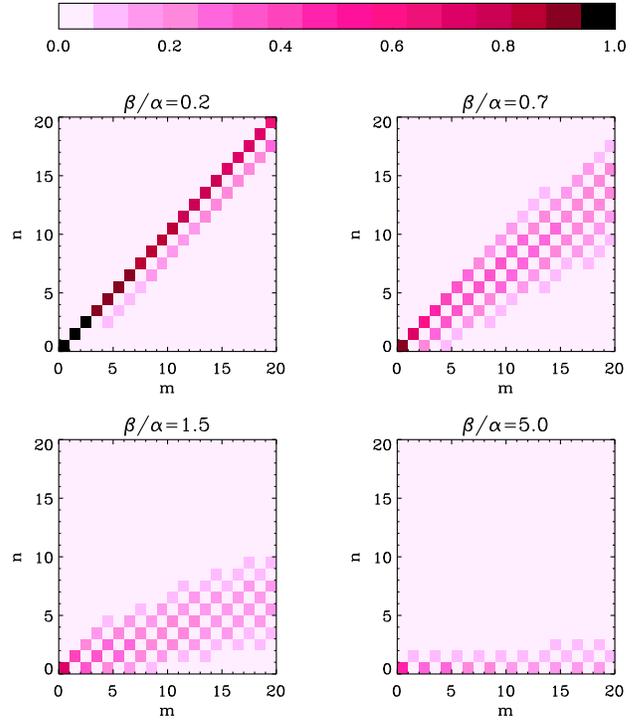,width=90mm}
\caption{Graphical representation of the smoothing matrix
$|G_{nm}|$ for different size $\beta$ of the smoothing kernel
(in units of the input shapelet scale $\alpha$). The horizontal
axis corresponds to the input (unsmoothed) mode $m$, while the vertical axis
shows the output (smoothed) mode $n$. For small smoothing
scales ($\beta \rightarrow 0$) the smoothing matrix approaches
the identity matrix (upper-left panel). For large smoothing scales
($\beta \rightarrow \infty$), it approaches a projection onto
the $n=0$ mode (lower-right panel). For intermediate values,
it corresponds to a projection onto lower order modes (upper-right
and lower-left panels).}
\label{fig:gmatrix}
\end{figure}

Figure~\ref{fig:smooth} shows how this analytic formula can be used to
efficiently smooth a 2-dimensional image (see discussion in
\S\ref{convolution_2d} below).  An intuitive feeling for the effect of
convolution on the shapelet coefficients can be obtained from
Figure~\ref{fig:gmatrix}, which graphically shows the smoothing matrix
$G_{nm}(\alpha,\beta,\gamma=(\alpha^2+\beta^2)^{\frac{1}{2}})$ for
different values of the smoothing scale $\beta$. As expected, the
smoothing matrix approaches the identity matrix in the limit of
vanishing smoothing scale ($\beta \rightarrow 0$). On the other hand,
for very large smoothing kernels ($\beta \rightarrow \infty$) it
reduces to a projection of all the input modes $m$ onto the $n=0$
output mode. For intermediate scales, the smoothing matrix takes the
form of a band which rotates from the vertical to the horizontal as
the smoothing scale $\beta$ is increased. Smoothing thus corresponds
to a projection of the input modes into output modes of smaller order.
The high-order modes indeed have oscillations on small scales
and are thus gradually lost when we increase the smoothing scale
$\beta$.

\subsection{Convolution in 2-Dimensions}
\label{convolution_2d}
Let us now consider the convolution of two 2-dimensional functions,
such as
\begin{equation}
h({\mathbf x})=(f*g)({\mathbf x})=\int d^{2}x~f({\mathbf x}-{\mathbf
x}') g({\mathbf x}').
\end{equation}
We again first decompose each function into shapelet coefficients
$f_{\mathbf n} \equiv \langle {\mathbf n};\alpha | f \rangle$,
$g_{\mathbf n} \equiv \langle {\mathbf n};\beta | g \rangle$, and
$h_{\mathbf n} \equiv \langle {\mathbf n};\gamma | h \rangle$ with
shapelet scales $\alpha$, $\beta$ and $\gamma$ respectively, and where
${\mathbf n}=(n_{1},n_{2})$ as before.  Because convolution is
bilinear we can again relate the convolved to the unconvolved
coefficients by
\begin{equation}
h_{\mathbf n} = \sum_{{\mathbf m},{\mathbf l}} 
  C_{{\mathbf n},{\mathbf m},{\mathbf l}} f_{\mathbf m} g_{\mathbf l}
\end{equation}
where $C_{{\mathbf n},{\mathbf m},{\mathbf l}}(\gamma,\alpha,\beta)$
is the 2-dimensional convolution tensor. From the separability of the
2-dimensional basis functions (see Eq.~[\ref{eq:phi_n1n2}), it is easy
to show that this tensor is equal to
\begin{equation}
C_{{\mathbf n},{\mathbf m},{\mathbf l}}(\gamma,\alpha,\beta) =
C_{n_{1},m_{1},l_{1}}(\gamma,\alpha,\beta)
C_{n_{2},m_{2},l_{2}}(\gamma,\alpha,\beta),
\end{equation}
where the tensors appearing on the right-hand side are the
1-dimensional convolution tensor defined in Equation~(\ref{eq:c_nml}).

We can also consider the special case of smoothing with a
2-dimensional gaussian. In this case, $g({\mathbf
x})=(2\pi\beta^2)^{-1} e^{-\frac{x^2}{2\beta^2}}$, which is normalised so
that $\int d^{2}x~g({\mathbf x})=1$. The smoothed coefficients are
then given by
\begin{equation}
h_{\mathbf n} = \sum_{{\mathbf m}} G_{{\mathbf n},{\mathbf m}}
f_{{\mathbf m}},
\end{equation}
where $G_{{\mathbf n},{\mathbf m}}(\gamma,\alpha,\beta)$ is the
2-dimensional smoothing matrix. It is again easy to show that it
is equal to
\begin{equation}
G_{{\mathbf n},{\mathbf m}}(\gamma,\alpha,\beta)=
G_{n_1,m_1}(\gamma,\alpha,\beta)
G_{n_2,m_2}(\gamma,\alpha,\beta),
\end{equation}
in terms of the 1-dimensional smoothing matrix defined in
\S\ref{smoothing_1d}. With the natural choice
$\gamma^{2}=\alpha^2+\beta^2$, it can be evaluated using 
Equation~(\ref{eq:g_nm}).

\begin{figure}
\psfig{figure=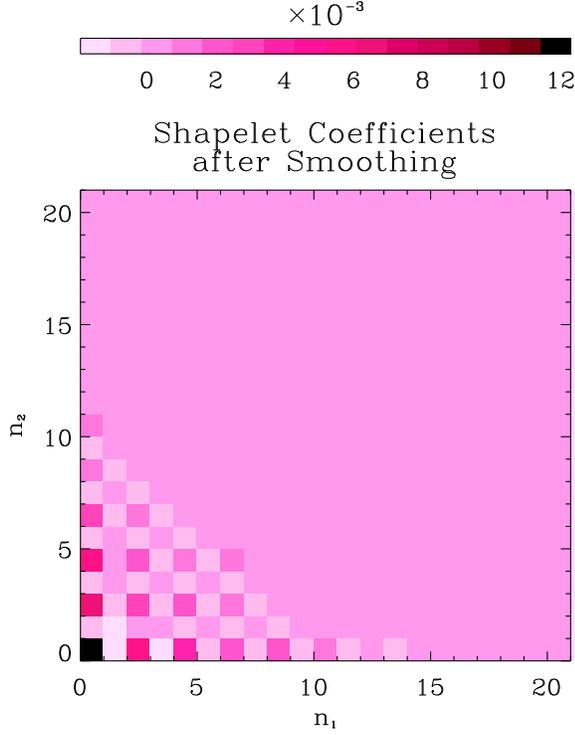,width=90mm}
\caption{Shapelet coefficients of the same galaxy (Figure~\ref{fig:hst})
after smoothing with a gaussian kernel. For clarity, the standard
deviation of the kernel was increased to $\beta=4$ pixels.  By
comparing this distribution with that before smoothing
(Figure~\ref{fig:hst_coef}), it is easy to see how convolution amounts
to a projection onto the lower order shapelet states.}
\label{fig:hst_coef_smooth}
\end{figure}

Figure~\ref{fig:smooth} shows the how the galaxy image of
Figure~\ref{fig:hst} can be smoothed using our shapelet method. The
resulting image is indistinguishable from that derived using ordinary
convolution in real-space (also shown). The shapelet method is however
computationally very efficient when the coefficient matrix is sparse
as is the case here (see Figure~\ref{fig:hst_coef}). The effect of
smoothing on the shapelet coefficients of this galaxy can be seen in
Figure~\ref{fig:hst_coef_smooth}. For clarity, the smoothing scale was
enhanced to $\beta=4$ pixels. Clearly, convolution amounts to a
projection onto the lower order states, as discussed in
\S\ref{smoothing_1d}.
 
\section{Polar Shapelets}
\label{two_d_polar}

The cartesian basis functions discussed above are separable in the
cartesian coordinates $x_{1}$ and $x_{2}$. It is also useful to
construct basis functions which are separable in the polar coordinates
$x$ and $\varphi$. These are eigenstates of the Hamiltonian
$\hat{H}$ and of the angular momentum $\hat{L}$ simultaneously, and
thus have a number of convenient features.  In this section, we show
how they can be constructed and study some of their properties.

\subsection{Raising and Lowering Operators}
To construct the polar basis functions, we first define the left and
right lowering operators as (see eg. Cohen-Tannoudji et al. 1977)
\begin{equation}
\label{eq:alar}
\hat{a}_{l} = \frac{1}{\sqrt{2}} \left( \hat{a}_{1} + i \hat{a}_{2}
\right),~~
\hat{a}_{r} = \frac{1}{\sqrt{2}} \left( \hat{a}_{1} - i \hat{a}_{2}
\right). 
\end{equation}
The associated raising operators are $\hat{a}_{l}^{\dagger}$ and
$\hat{a}_{r}^{\dagger}$, respectively. The only non-vanishing
commutators between these operators are
\begin{equation}
[\hat{a}_{l},\hat{a}_{l}^{\dagger}] =
[\hat{a}_{r},\hat{a}_{r}^{\dagger}]=1.
\end{equation}
The Hamiltonian (Eq.~[\ref{eq:hamiltonian_2d}]) and angular momentum
(Eq.~[\ref{eq:l}]) operators for the 2-dimensional QHO can be then be
written as
\begin{equation}
\label{eq:hl_lr}
\hat{H} = \hat{N}_{r}+\hat{N}_{l}+1,~~~\hat{L}=\hat{N}_{r}-\hat{N_{l}},
\end{equation}
where the 
left-handed and right-handed number operators are naturally
defined as
\begin{equation}
\hat{N}_{l} = \hat{a}_{l}^{\dagger} \hat{a}_{l}, ~~~
\hat{N}_{r} = \hat{a}_{r}^{\dagger} \hat{a}_{r}.
\end{equation}
The operators $\hat{a}_{l}^{\dagger}$, $\hat{a}_{r}^{\dagger}$,
$\hat{a}_{l}$, and $\hat{a}_{r}$ can thus be thought of as creating
and destroying left- and right-handed quanta.

\subsection{Angular Momentum States}
Since the operators $\hat{N}_{r}$ and $\hat{N_{l}}$ form a complete
set of commuting observables, their eigenstates $\vert n_{l},n_{r}
\rangle$ provide a complete basis for our function space. These states
are defined $\hat{N}_{l}\vert n_{l},n_{r} \rangle = n_{l} \vert n_{l},n_{r}
\rangle$, and similarly for $\hat{N}_{r}$, for $n_{l}$ $n_{r}$
non-negative integers. They can be constructed by applying the raising
operators several times on the ground state $\vert n_{1}=0,n_{2}=0 \rangle
= \vert n_{l}=0,n_{r}=0 \rangle \equiv \vert 0,0 \rangle$, as
\begin{equation}
\label{eq:nrnl}
\vert n_{r},n_{l} \rangle = \frac{(\hat{a}_{r}^{\dagger})^{n_{r}}
(\hat{a}_{l}^{\dagger})^{n_{l}}}{\sqrt{n_{r}! n_{l}!}} \vert 0,0 \rangle.
\end{equation}
From Equation~(\ref{eq:hl_lr}), it is easy to see that
\begin{equation}
\hat{H} \vert n_{l}, n_{r} \rangle = (n_{r}+n_{l}) \vert n_{l}, n_{r} \rangle,~~
\hat{L} \vert n_{l}, n_{r} \rangle = (n_{r}-n_{l}) \vert n_{l}, n_{r} \rangle.
\end{equation}
We can therefore relabel these states in terms of their energy and
angular momentum quantum numbers, $n=n_{r}+n_{l}$ and $m=n_{r}-n_{l}$,
as 
\begin{equation}
\vert n,m \rangle = \vert n_{l}=\frac{1}{2} (n-m),
n_{r}=\frac{1}{2} (n+m) \rangle.
\end{equation}
The angular momentum quantum number takes on the $n+1$ values given by
$m=-n,-n+2,\ldots,n-2,n$.

\subsection{Basis Functions}
Using the $x$-representation of $\hat{a}_{l}^{\dagger}$ and
$\hat{a}_{r}^{\dagger}$, one can show that the basis functions
$\chi_{n_l,n_r}(x,\varphi) \equiv \langle x \vert n_{l},n_{r} \rangle$ for
the angular momentum states are given by
\begin{equation}
\chi_{n_l,n_r}(x,\varphi) = \left[ \pi n_{l}! n_{r}!
\right]^{-\frac{1}{2}} H_{n_{l},n_{r}}(x) e^{-x^{2}/2}
e^{i(n_{r}-n_{l})\varphi},
\end{equation}
where $H_{n_{k},n_{r}}(x)$ are polynomials, which we call `polar
Hermite polynomials'. They can be computed by noting that
$H_{0,0}(x)=1$ and by using the recursion relation
\begin{equation}
\frac{l-k}{x} H_{k,l}(x) = l H_{k,l-1} - k H_{k-1,l}.
\end{equation}
The diagonal polynomials can be computed using
\begin{equation}
H_{kk} = H_{k+1,k-1} - x^{-1} H_{k,k-1}.
\end{equation}
The first few polar Hermite polynomials are listed in
Table~\ref{tab:hpolar}. They have a number of useful properties. In
particular, they are symmetric, i.e. $H_{k,l}=H_{l,k}$ and their
derivative obey
\begin{eqnarray}
H'_{k,l}(x) & = & k H_{k-1,l}(x) + l H_{k,l-1}(x) \nonumber \\
   & = &  2x H_{k,l}(x)-H_{k+1,l}-H_{k,l+1}.
\end{eqnarray}

\begin{figure}
\psfig{figure=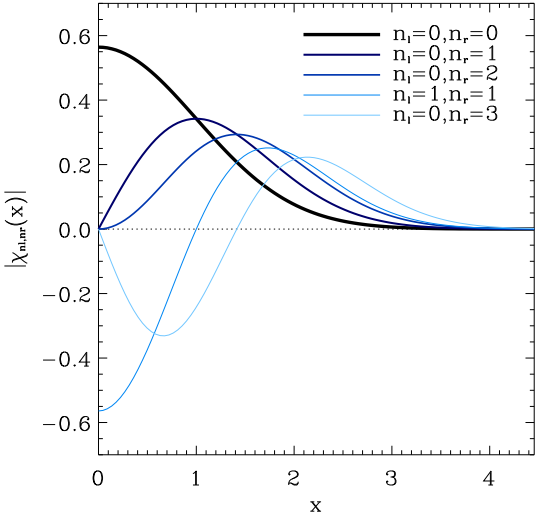,width=80mm}
\caption{Radial dependence of the first few polar basis functions
$|\chi_{n_l,n_r}(x)|$.}
\label{fig:polar}
\end{figure}

Dimensional polar basis functions can be constructed as
\begin{equation}
A_{n_l,n_r}(x,\varphi;\beta) = \beta^{-1} \chi_{n_l,n_r}(\beta^{-1}x,\varphi).
\end{equation}
It is easy to check that these are orthonormal, i.e. that
\begin{eqnarray}
\int_0^{2\pi} d\varphi \int_{0}^{\infty} dx~ x
A_{n_l,n_r}(x,\varphi;\beta) A_{n_l',n_r'}(x,\varphi;\beta) & &
\nonumber
\end{eqnarray}
\begin{equation}
= \langle {n_l,n_r;\beta} | {n_l',n_r';\beta} \rangle
= \delta_{n_l,n_l'} \delta_{n_r,n_r'}.
\end{equation}
The radial dependence $|\chi_{n_l,n_r}(x)|$ of the first few
polar shapelet functions is shown in Figure~\ref{fig:polar}.

\begin{table}
\centering 
\begin{minipage}{140mm} 
\caption{First few polar Hermite polynomials}
\label{tab:hpolar} 
\begin{tabular}{l} 
\hline
$H_{0,0}(x)=1$ \\
$H_{0,1}(x)=H_{1,0}(x)=x$ \\
$H_{0,2}(x)=H_{2,0}(x)=x^{2}$\\
$H_{1,1}(x)=x^{2}-1$ \\
$H_{0,3}(x)=H_{3,0}(x)=x^{3}$\\
$H_{1,2}(x)=H_{2,1}(x)=x^{3}-2x$\\
$H_{0,4}(x)=H_{4,0}(x)=x^{4}$\\
$H_{1,3}(x)=H_{3,1}(x)=x^{4}-3x^{2}$\\
$H_{2,2}(x)=x^{4}-4x^{2}+2$\\
\hline
\end{tabular}
\end{minipage}
\end{table}

\subsection{Relation to Cartesian States}
\label{pol_cart}
It is useful to relate the angular momentum states $\vert n,m \rangle$
to the cartesian states $\vert n_{1}, n_{2} \rangle$. Using
Equations~(\ref{eq:alar}) and (\ref{eq:nrnl}) along with the binomial
expansion, one can show that the transformation matrix between these
two bases is given by
\begin{eqnarray}
\langle n_{1},n_{2} \vert n_{l}, n_{r} \rangle & = &
 2^{-\frac{1}{2}(n_r+n_l)} i^{n_r-n_l} 
\left[ \frac{n_1 ! n_2 !}{n_r ! n_l !} \right]^{\frac{1}{2}} \nonumber
 \delta_{n_1+n_2 , n_r + n_l}
 \\
 & & \sum_{n_r'=0}^{n_r} \sum_{n_l'=0}^{n_l} i^{n_l'-n_r'}
\left( \begin{array}{c} n_{r} \\ n_{r}' \end{array} \right)
\left( \begin{array}{c} n_{l} \\ n_{l}' \end{array} \right)
\delta_{n_r'+n_l' , n_1}.
\end{eqnarray}
This shows that only states with $n_1+n_2=n_r+n_l$ are mixed. The
first few $\vert n,m \rangle$ states are given in terms of $\vert n_1
n_2 \rangle$ states in Table~\ref{tab:nm_states}.

\begin{table}
\centering 
\begin{minipage}{80mm} 
\caption{Angular momentum states $\vert n,m \rangle$ in terms of the
cartesian states $\vert n_1,n_2 \rangle$.}
\label{tab:nm_states} 
\begin{tabular}{l} 
\hline
$\vert n=0,m=0 \rangle = \vert 0,0 \rangle$ \\
$\vert n=1,m=1 \rangle = \frac{1}{\sqrt{2}} \left[ \vert 1,0 \rangle
  + i \vert 0,1 \rangle \right]$ \\
$\vert n=1,m=-1 \rangle = \frac{1}{\sqrt{2}} \left[ \vert 1,0 \rangle
  - i \vert 0,1 \rangle \right]$ \\
$\vert n=2,m=2 \rangle = \frac{1}{2} \left[ \vert 2,0 \rangle
  + i \sqrt{2} \vert 1,1 \rangle - \vert 0,2 \rangle \right]$ \\
$\vert n=2,m=0 \rangle = \frac{1}{\sqrt{2}} \left[ \vert 2,0 \rangle
  + \vert 0,2 \rangle \right]$ \\
$\vert n=2,m=-2 \rangle = \frac{1}{2} \left[ \vert 2,0 \rangle
  - i \sqrt{2} \vert 1,1 \rangle - \vert 0,2 \rangle \right]$ \\
\hline
\end{tabular}
\end{minipage}
\end{table}

\subsection{Properties}
Because the polar shapelet states are eigenstates of the angular
momentum, they have simple rotational properties. Indeed, under
a finite rotation by an angle $\rho$, the polar states transform as
\begin{equation}
|n_r,n_l\rangle'=e^{-i \rho \hat{L}} |n_r,n_l\rangle = 
e^{-i \rho (n_r-n_l)} |n_r,n_l\rangle,
\end{equation}
where we have used the exponentiation (Eq.~[\ref{eq:rot_finite}]) of
the rotation generator $\hat{R}=-i\hat{L}$ (Eq.~[\ref{eq:generators}])
to operate a finite rotation. In this basis, finite rotations thus
corresponds only to a phase factor. 

It is therefore a simple matter to rotate an arbitrary function
$f({\mathbf x})$. First, we decompose it into polar shapelet
coefficients $f_{n_r,n_l}=\langle n_r,n_l;\beta | f \rangle$, with an
appropriate shapelet scale $\beta$. The coefficients 
$f_{n_r,n_l}'=\langle n_r,n_l;\beta | f' \rangle$ of the
rotated function $f'({\mathbf x})$ are then given simply by
\begin{equation}
f'_{n_r,n_l} = e^{-i (n_r-n_l) \rho} f_{n_r,n_l}.
\end{equation}

By contrast, operating a finite rotation in the cartesian basis
requires an infinite number of applications of the $\hat{R}$ operator
(see Eq.~[\ref{eq:rot_finite}]) and is thus impractical.  On the
other hand, convolutions do not have simple analytical expressions in
the polar basis, as they do in the cartesian basis (see
\S\ref{convolution} and Paper II). The results of \S\ref{pol_cart},
can thus be conveniently used to convert from one basis to the other,
depending on the operation to be performed.

\section{Applications}
\label{applications}
Now that we have developed the main formalism for shapelets, we
illustrate the method using images from the Hubble Space Telescope
(HST). We also discuss several direct applications of shapelets.

\subsection{Example with HST images}
As an example, we apply the shapelet decomposition method to images of
galaxies found in the Hubble Deep Field (HDF; Williams et al. 1996),
the deepest images observed with the Hubble Space telescope
(HST). Figure~\ref{fig:hst} shows the original $61 \times 61$ image
$f({\mathbf x})$ of one such galaxy (upper-left hand panel).  Using
Equation~(\ref{eq:decompose_2d}), we first compute the shapelet
coefficients $f_{\mathbf n}$ of the galaxy with a shapelet scale
$\beta=4$ pixels.  We then reconstruct the image using
Equation~(\ref{eq:coefs_2d}) including coefficients up to a maximum
order $n_{1}+n_{2} \le n_{\rm max}$. The resulting reconstructed
images are shown on the figure for different values of $n_{\rm
max}$. As $n_{\rm max}$ is increased, more small scale and large scale
features are recovered, as expected from the properties of our basis
functions (see \S\ref{th_min_max_1d}). For $n_{\rm max}=20$, the reconstructed
image is almost indistinguishable from the original.

Figure~\ref{fig:hst_coef} gives a graphical representation of the
shapelet coefficient matrix $f_{n_{1},n_{2}}$ for this image. This can
be thought as the representation of the galaxy in `shapelet space'. As
is apparent on the figure, the coefficient matrix is rather sparse.
The fact that the odd coefficients are small results from the fact
that the shapelet center was chosen to be close to the centroid of the
galaxy.  The coefficients are negligibly small beyond $n_{1}+n_{2} \ga
15$, thus explaining why we obtain a virtually full reconstruction
with $n_{\rm max}=20$.

Because of the sparse nature of the coefficient matrix, we can hope to
recover the image from only the first few largest coefficients.
Figure~\ref{fig:compress} shows the reconstruction of the galaxy of
Figure~\ref{fig:hst} and of two other HDF galaxies, by keeping only
the $N_{\rm cof}$ coefficients with largest absolute value
$|f_{\mathbf n}|$. As can be seen from the top two panels, the galaxy
image can be faithfully recovered with the top $N_{\rm cof}=60$
coefficients, yielding a compression factor $N_{\rm pix}/N_{\rm cof}$
of 62 compared to the original image which contained $N_{\rm
pix}=61\times61=3721$ pixels. (The bookeeping required to keep track
of selected coefficients only requires 1 bit per coefficient, and thus
results only in a small overhead relatively to this compression
factor). For the other two galaxies shown in the middle and bottom
rows, compression factors of about 40-90 are achieved. Note that the
galaxy in the bottom row has a simpler structure and thus affords
more compression.

\subsection{Catalogue Archival}
We have seen above that the first few shapelet coefficients capture
most of the structure of galaxy images and thus allow strong data
compressions. This can be very useful for upcoming and future large
galaxy surveys such as the Sloan Digital Sky Survey (SDSS; Gunn \&
Weinberg 1995), or that derived from the planned SNAP mission
(Perlmutter et al. 2001). One can imagine storing the first few
shapelet coefficients in the catalog, thus both saving storage and
conveying compactly the shape information of each each galaxy. The flux,
centroid, major and minor axes and position angle of each galaxy could
then be computed from their shapelet coefficients directly (as
described in \S\ref{astrometry}), thus avoiding the need to consider several
definitions of magnitudes.  Since galaxy shapes in different wavelengths
are strongly correlated, the treatment of multi-colour data could be
done efficiently by decomposing differences of the images in different
pass-bands and again keeping the largest coefficients. The resulting
catalogue could then also be useful for study of galaxy morphology and
classification in shapelet space.

\subsection{Modeling the Point-Spread Function}
\label{PSF_modeling}
Several astronomical techniques (eg. high-precision astrometry and
photometry, microlensing, weak lensing, supernova searches, etc)
require correction for the Point-Spread Function (PSF) of the
telescope across an image. For instance, Alard \& Lupton (1998) have
developed a technique to take the difference of two images convolved
with a spatially varying PSF. Shapelets provide a convenient
correction method for the PSF: the PSF shape can be measured at
different positions in the field using bright stars and then decomposed
into shapelet coefficients; a 2-dimensional polynomial fit for each
shapelet coefficient as a function of position can then be performed
to derive a model of the PSF shape at any point (Cf. Alard \& Lupton
1998 and Kaiser 2000 who used this approach with other sets of basis
functions). The convolution matrix can then be inverted to compute the
shapelet coefficients of objects prior to convolution. As discussed
in \S\ref{convolution}, convolution amounts to a projection onto lower
shapelet order. As a result only low order coefficients can be
recovered. Another approach consist of fitting the deconvolved
shapelet coefficients convolved to the PSF model to the data
(Cf. Kuijken 2000). The analytical properties of shapelets under
convolution (see \S\ref{convolution} and Paper II) greatly facilitate
and clarify the procedure. A detailed study of deconvolution using
shapelets will be presented in Paper II.

\subsection{Gravitational Lensing}
Gravitational Lensing is a powerful method to directly probe the mass
of astrophysical objects. In particular, the weak coherent distortions
induced by lensing on the images of background galaxies provide a
direct measure of the distribution of mass in the Universe. This weak
lensing method is now routinely used to study galaxy clusters, and has
recently been detected in the field (see reviews by Mellier 1999;
Bartelmann \& Schneider; Mellier et al. 2000). Because the lensing effect is
only of a few percent on large scales, a precise method for measuring
the shear is required. The original methods of Bonnet \& Mellier
(1995) and of Kaiser, Squires \& Broadhurst (KSB; 1995) are not
sufficiently accurate and stable for the upcoming weak lensing
surveys. Thus, several new methods have been proposed (Kuijken 1999;
Rhodes, Refregier \& Groth 2000; Kaiser 2000). As we briefly describe
below, the remarkable properties of our basis functions, make
shapelets particularly well suited for providing the basis of a new
method.

Let us consider a galaxy with an unlensed intensity $f({\mathbf x})$.
We have shown in \S\ref{transformations} that under the action of a
weak shear $\gamma_{i}$, the lensed intensity is
\begin{equation}
f' \simeq (1+\gamma_{i} \hat{S}_{i}) f.
\end{equation}
After decomposing these intensities into our basis functions
$B_{{\mathbf n}}({\mathbf x},\beta)$
(Eq.~[\ref{eq:decompose_2d}]), this becomes a relation between the lensed and
the unlensed coefficients given by
\begin{equation}
f'_{{\mathbf n}} = (\delta_{nm}+\gamma_{i} S_{i{\mathbf mn}}) f_{{\mathbf m}},
\end{equation}
where $S_{i{\mathbf mn}} \equiv \langle {\mathbf m} | \hat{S}_{1} |
{\mathbf n} \rangle$ is the shear generator matrix given in Equation
(\ref{eq:generators}). The goal for weak lensing is to estimate the
shear from the shapes of an ensemble of galaxies which are assumed to
be randomly oriented prior to lensing. In the widely used KSB method,
this is done by considering the effect of lensing on the
Gaussian-weighted quadrupole moments of the galaxy images. These are
exactly equal to the $n_1+n_2=2$ coefficients in our shapelet
decomposition. In as sense, our method thus generalises this
approach and capture all the available shape information of
galaxies. Because the shear matrix is simple in our cartesian basis,
we can then construct an estimator for the shear by comparing the
distribution of the lensed shapelet coefficients $f_{\mathbf n}'$ to
that of a training set $f_{\mathbf n}$ for which lensing is known to
be negligible. This can be done either by constructing a linear shear
estimator from the observed coefficients or by using a Maximum
Likelihood technique. In Paper II, we follow the first approach and
show that shapelets can be used to derive precise shear recovery in
realistic simulations of deep optical images (see Bacon et al. 2001).

As Luppino \& Kaiser (1997) discussed, the shear acts, in practice,
after the smearing produced by the PSF. To account for this, we can
first model the PSF across the field, as described in
\S\ref{PSF_modeling}. Then one of the methods mentioned in this
section can be used to derive the deconvolved coefficients from the
observed convolved ones. Again, a detailed study of this deconvolution
method and its impact in weak lensing measurement will be presented in
Paper II.

Shapelets can also be used for strong lensing applications, such as
the modeling of cluster or galaxy potentials using multiple images and
giant arcs. In \S\ref{transformations}, we concentrated mainly on
first order distortions, but also mentioned that distortions of
arbitrary amplitudes can be derived by exponentiating the shear and
convergence operators (see Eq.~[\ref{eq:rot_finite}]). An equivalent
method to compute the effect of large distortions on the shapelet
coefficients is to use the analytical expression for rescaling of
Appendix~\ref{rescaling}. One can then model the shape of the lensed
object using shapelets and explore a large class of lens models
efficiently by computing the lensed image coefficients analytically.
Another possibility is to also model the gravitational potential of
the lens using shapelets.

\subsection{Deprojection}
Another important problem in astronomy is that of deprojection.  For
instance, the 2-dimensional images of galaxies and clusters of galaxies
observed on the sky are projections of their 3-dimensional
distributions. One can hope to reconstruct the 3-dimensional
distribution of these systems by combining observations at different
wavelengths. These indeed probe different physical processes, and
therefore correspond to different weighting along the line of
sight. Here, we show how shapelets can be used to solve this problem.

To so so, we consider the simple, yet practical example of a cluster
of galaxies observed both through its X-ray emission (see eg. Sarazin
1988 for a review) and Sunyaev-Zel'dovich (SZ) effect (Sunyaev \&
Zel'dovich 1972; see Birkinshaw 1999, for a review). Cluster
deprojection is a long standing problem in astrophysics, and has been
studied by several groups (see the recent solution by Dor\'{e} et
al. 2001, and references therein). Here, we assume, for simplicity,
that the cluster gas is isothermal. The X-ray emissivity of the
cluster can then be written as (eg. Sarazin 1988)
\begin{equation}
X(x,y) \simeq X_{0} \int dz~\rho^{2}(x,y,z),
\end{equation}
where $\rho$ is the 3-dimensional electron density of the gas, $z$ is
the line-of-sight coordinate, and $X_{0}$ is a constant which depends
on the wavelength of observation, the gas temperature, and the
distance to the cluster. The comptonisation parameter $Y(x,y)$ from
the SZ effect can be observed as temperature anisotropies of the
Cosmic Microwave Background (see review by Birkinshaw 1999). It is
proportional to the electron pressure integrated along the line of
sight, and thus be written, for an isothermal cluster as
\begin{equation}
Y(x,y) \simeq Y_{0} \int dz~\rho(z,y,z),
\end{equation}
where $Y_{0}$ is a constant which again depends on the wavelength of
observation, the gas temperature, and distance to the cluster. Our goal
is to reconstruct the three-dimensional gas density $\rho(x,y,z)$ from
measurements of $X(x,y)$ and $Y(x,y)$.

For this purpose, let us decompose these two observed images into
2-dimensional shapelets as $X(x,y) = \sum_{n_{1}n_{2}} X_{n_{1}n_{2}}
B_{n_{1}n_{2}}(x,y)$,~and~$Y(x,y) = \sum_{n_{1}n_{2}} Y_{n_{1}n_{2}}
B_{n_{1}n_{2}}(x,y)$. We choose the same shapelet scale $\beta$ for
$X$, $Y$ and $\rho$ and thus drop it to simplify the notation.  In
analogy with the discussion in \S\ref{2d_cartesian_def}, we can also
define 3-dimensional basis functions $B_{n_{1}n_{2}n_{3}}(x,y,z)
\equiv B_{n_{1}}(x) B_{n_{2}}(y) B_{n_{3}}(z)$ as products of three
1-dimensional shapelets. This allows us to decompose the
3-dimensional gas density distribution as
\begin{equation}
\rho(x,y,z) = \sum_{n_{1},n_{2},n_{3}} \rho_{n_{1}n_{2}n_{3}}
B_{n_{1}n_{2}n_{3}}(x,y,z).
\end{equation}
Using the properties of the shapelet basis functions, it is then easy
to show that the shapelet coefficients for the X-ray emissivity 
can be written as
\begin{equation}
X_{n_{1}n_{2}} = X_{0} 
\sum_{{\mathbf m},{\mathbf m}'}
  B_{n_{1}m_{1}m_{1}'}^{(3)} B_{n_{2}m_{2}m_{2}'}^{(3)} \delta_{m_{3},m_{3}'}
  \rho_{{\mathbf m}} \rho_{{\mathbf m}'},
\end{equation}
where $B^{(3)}_{nml}$ is the ubiquitous 3-product integral defined in
Equation~(\ref{eq:b3}), and ${\mathbf m} \equiv (m_{1},m_{2},m_{3})$ in
this context. Similarly, the coefficients for the comptonisation
parameter can be written as
\begin{equation}
Y_{n_{1}n_{2}} = Y_{0} 
\sum_{n_{3}} \langle 1 | n_{3} \rangle \rho_{{\mathbf n}},
\end{equation}
where $\langle 1 | n \rangle$ is the integral defined in
Equation~(\ref{eq:1n}). The direct approach, which consists in solving
these two equations of the desired coefficients $\rho_{{\mathbf n}}$,
is probably difficult in practice. A more convenient approach is to
derive an estimate for $\rho_{{\mathbf n}}$ by $\chi^{2}$-fitting
these coefficients to the observables $X_{n_{1}n_{2}}$ and
$Y_{n_{1}n_{2}}$ taking into account the noise in each
measurement. The $\chi^{2}$ procedure also produces the covariance
matrix for the coefficients $\rho_{{\mathbf n}}$, and thus allows us
to study any degeneracy present in the deprojection.  This is greatly
facilitated in practice by the analytic forms for $\langle 1 | n
\rangle$ (Eq.~[\ref{eq:1n}]) and for $B^{(3)}_{nml}$ (see Paper II),
and the fact that the fitted model is linear in its output parameters
$\rho_{{\mathbf n}}$.

Note that our method is fully general and does not assume that the
cluster distribution has any specific form. In particular, it could be
particularly useful if the SZ observations are performed with an
interferometer as is the case for recent measurements (see Carlstrom
et al. 1999, and reference therein). In this case, the interferometer
yields a measurement of the Fourier transform of $Y(x,y)$, and can
thus make use of the dual properties of our shapelet functions under
Fourier transforms (Eq.~[\ref{eq:B_tilde}]). A more thorough study of the
deprojection using shapelets is left to future work. A study of the
use of shapelets for reconstructing images with interferometers will
be presented in Chang \& Refregier (2001).

\section{Conclusions}
\label{conclusion}
We have described and developed a new method for analysing images. It
is based on the decomposition of the objects in the image into a
series of basis functions of different shapes, or `shapelets'.  The
method is fully linear and uses a number of powerful properties of the
basis functions. In particular, we showed that Hermite basis functions
have simple analytic properties under convolution, noise, rotations,
distortions, and rescaling. These functions are eigenfunctions of the
QHO and thus allow us to use the formalism developed for this
problem. For instance, we showed that transformations such as
translations, rotations, shears and dilatations can be expressed as
simple combinations of the raising and lowering operators.  Another
remarkable property of these functions is that they are (up to a
rescaling) their own Fourier transforms. This is a unique property,
which stems from the special symmetry of the QHO Hamiltonian. We
derived analytical expressions for the flux, centroid and radius of
the object, from its shapelet coefficients. We also constructed polar
shapelets which give the explicit rotational properties of the object.

It is interesting to compare our method to the wavelet method (see
review by Stark, Murtagh \& Bijaoui 1998). In this latter method, the
image is decomposed into a sum of basis functions located on a grid
across the image. The basis functions are taken to have a range of
sizes, but have all the same shape. Wavelets are thus ideal to
decompose an image into different scales, which can then be analysed
separately. Our method, on the other hand models the image as a
collection of discrete objects of arbitrary shapes and sizes. It is
therefore particularly well adapted to the treatment of astronomical
images, which are typically composed of a superposition of compact
disjoint objects. The two methods can thus be thought as
complementary. For instance, one can use wavelets to remove
large-scale background variations, and to search and detect objects in
the image. The resulting object catalog can then be
used as the input to the shapelet method, which will then characterise
the shape of each object in detail.

Our method potentially has a wide range of applications. It can be
viewed as a new representation of images which makes object shapes
easy to study and modify. For instance, we applied our method to
galaxy images found in the HDF and showed how they could be well
represented with a small number of shapelet coefficients. This can be
used to compress galaxy images by a factor of 40-90, and could thus
have important applications for galaxy archival. We also discussed
several direct applications of shapelets to measurements of
gravitational lensing, and the problems of de-projection and PSF
correction.  Other applications to be explored are that of
multi-colour shapelets and of the study of galaxy morphology and
classification using shapelets.  Our original motivation for
developing this method was to find a robust and precise method to
measure weak lensing distortions in the presence of a PSF. The
application of shapelets to this problem and to the general problem of
deconvolution will be presented in detail in Paper II. The application
of shapelets to image reconstructions from interferometric observations
will be presented in Chang \& Refregier 2001.

\section*{Acknowledgments}
The author is indebted to David Bacon, Tzu-Ching Chang and R. Chitra for
useful and stimulating discussions during the development of this method.
He is supported by the EEC TMR network on Gravitational Lensing and
by a Wolfson College Research Fellowship.

\appendix

\section{Rescaling}
\label{rescaling}
In this appendix, we show how we can easily operate a change of scale
$\beta$ for the decomposition of a function in 1-dimension. This is
convenient for finding the optimal scale $\beta$ for a given
function. In addition, such a change of scale occurs when a
2-dimensional image is distorted by gravitational lensing.

Let us consider a function $f(x)=\langle x | f \rangle$ which we
decompose (as in Eq.~[\ref{eq:decompose}]) into two sets of basis
functions with scales $\beta_{1}$ and $\beta_{2}$ as
\begin{equation}
f(x) = 
\sum_{n=0}^{\infty} \langle n; \beta_{1} | f \rangle~
B_{n}(x;\beta_{1}) =
\sum_{n=0}^{\infty} \langle n; \beta_{2} | f \rangle~ B_{n}(x;\beta_{2})
\end{equation} 
The coefficients in each basis are related by
\begin{equation}
\langle n_{1}; \beta_{1} | f \rangle = \sum_{n_{2}=0}^{\infty} 
\langle n_{1}; \beta_{1} | n_{2}; \beta_{2} \rangle 
\langle n_{2}; \beta_{2} | f \rangle.
\end{equation}
Using the generating function of Hermite polynomials, one can show
that the transformation matrix is given by
\begin{eqnarray}
\label{eq:beta1_beta2}
\langle n_{1}; \beta_{1} | n_{2}; \beta_{2} \rangle & = &
\sum_{l=0}^{\min(n_{1},n_{2})} 
 (-1)^{\frac{n_{2}-l}{2}} \frac{\left( n_{1}! n_{2}! \right)^{\frac{1}{2}}}
{\left(\frac{n_{1}-l}{2}\right)! \left(\frac{n_{2}-l}{2}\right)! l!}
\nonumber \\
 &  & \times \Pi(n_{1},n_{2},l) \left( \frac{b_{1}}{2} \right)^{\frac{n_{1}+n_{2}}{2}-l}
b_{2}^{l+\frac{1}{2}},
\end{eqnarray}
where 
\begin{equation}
b_{1} \equiv
\frac{\beta_{1}^{2}-\beta_{2}^{2}}{\beta_{1}^{2}+\beta_{2}^{2}},~~~
b_{2} \equiv
\frac{2\beta_{1}\beta_{2}}{\beta_{1}^{2}+\beta_{2}^{2}}
\end{equation}
and $\Pi(n_{1},n_{2},l)$ is equal to $1$ if $n_{1}$, $n_{2}$ and $l$
are all odd or all even, and is equal to 0 otherwise. In the limiting
case where $\beta_{1}=\beta_{2}$, the transformation matrix reduces to
$\langle n_{1}; \beta_{1} | n_{2}; \beta_{2} \rangle =
\delta_{n_{1},n_{2}}$, in agreement with the orthonormal properties of
the basis functions (Eq.~[\ref{eq:orthonorm}]).


\bsp
\label{lastpage}

\end{document}